# Thermo-optic tuning of directional infrared emissivity

Jae S. Hwang[1], Jin Xu[1] and Aaswath P. Raman[1,2*]

[1] Department of Materials Science and Engineering, University of California, Los Angeles, Los Angeles, CA 90095 USA

[2] California NanoSystems Institute, University of California, Los Angeles, Los Angeles, CA 90095 USA

*Corresponding Author: aaswath@ucla.edu

**Abstract**

Tuning the spatial extent of directional thermal emission across an arbitrary, and fixed spectral bandwidth is a fundamentally enabling capability for a range of emerging applications such as thermophotovoltaics, thermal imaging, and radiative cooling. However, previous experimental demonstrations were limited to narrow bandwidths, and the resonance frequency itself changed significantly as a function of the reconfigured directional response. Here, we demonstrate thermo-optic tuning of directional infrared emissivity using InAs-based gradient ENZ materials functioning as broadband directional thermal emitters whose angular selectivity can be modified via thermal free-carrier effects. We experimentally demonstrate two emitters achieving a 5° and 10° increase in the angular extent of their directional emissivity in the p-polarization across a prescribed, broad wavelength range of operation (12.5 to 15$\mu$m), for moderate temperatures below 400 K. Temperature-driven control of directional emissivity offers a new mode of post-fabrication control of radiative heat transfer that may in turn enable novel device functionalities.

**Introduction**

Thermal emission is generally assumed to be incoherent in both space and time, with its light emission toward the far-field being broadband and omnidirectional[1]. Deliberately constraining thermally generated light emission to specific angular channels across a broadband frequency range is a fundamentally enabling capability[2], which has stimulated an active field of research both theoretically[2,3] and experimentally[4-7]. In this context, nanophotonic structures with post-fabrication tunability have garnered significant attention and have been increasingly explored more recently. Reconfigurability of the spatial or temporal content of broadband far-field thermal radiation offers transformative potential across a broad range of applications such as thermal imaging[8-10], sensing[11-13], and energy conversion[14-16].

While significant progress has been made in tuning spectral properties using external stimuli[17-27] (e.g. electrical gating, optical pumping or temperature), dynamically controlling the angular response of thermal emission has proven to be more challenging[28]. Electrostatic modulation of the directionality of thermal emission has been experimentally explored in gap plasmon polariton based micro strip cavity modes[29-31] and Fabry-Perot resonance cavities supporting long-range delocalized modes[32]. However, the spectral bandwidth of directional emissivity was narrowband in nature and the resonance frequency itself changed significantly as a function of the peak emissivity angle of incidence. Alternatively, doped III-V semiconductor-based gradient epsilon-near-zero (ENZ) materials can be designed to behave as broadband directional thermal emitters[5]. Since semiconductor-based gradient ENZ photonic structures are driven by free carrier (electron or hole) concentrations, this approach holds great potential for enabling on-demand spatio-temporal control of emissivity through free-carrier effects. Recent work, for instance, has experimentally demonstrated that doped semiconductor-based gradient ENZ photonic structures can be used to realize broadband nonreciprocal thermal emitters by externally applying a static magnetic field in the Voigt geometry[33,34].

Thermal emission, however, is intrinsically a temperature dependent phenomenon[1]. Controlling thermal emission through temperature effects has been studied in the past, and semiconductors have been investigated in this context due to their temperature dependent free-carrier effects[35-37]. Modal refractive index shifts in semiconductor Mie resonators have been shown based on traditional thermo-optic effects[38-40]. Spectrally tunable thermal antennas were also demonstrated by thermally tuning the plasma frequency of III-V semiconductors[41]. However, prior work has not demonstrated the ability to tune the angular extent of a broadband directional emitter, and the spectral extent of the emitter was limited to narrow operational bandwidths centered at the tunable excitation frequency of the fundamental optical mode, with no distinct directionality.

Here, we explore the thermo-optic tuning of a broadband directional thermal emitter whose angular extent of "thermal beaming", and thus its angular selectivity, can be controlled via thermal free-carrier effects. We experimentally demonstrate this capability via semiconductor gradient ENZ photonic structures based on epitaxially grown graded doped InAs. We show this effect for photonic structures manifesting, by design, highly directional thermal emission, peaking at 75° and 65° respectively, across a broad bandwidth between 12.5 to 15$\mu$m. We show that the angular width of average emissivity > 0.4 over 2$\mu$m bandwidths can be gradually modulated by 5° and 10° for the respective structures within moderate temperature ranges (< 400 K), hence, achieving reconfigurable angular selectivity across a broad emissivity bandwidth. We describe the temperature-dependent optical response using theoretical and empirical models, showing that the angular range of high emissivity can be further reduced by 8° and 15° at lower temperatures (200 K) compared to the spectra at 375 K for two fabricated structures enabling tailored control over the directional range of high emissivity and angular selectivity, while maintaing the spectral response over a prescribed emissivity bandwidth by design.

**Results and Discussion**

To explore thermal control of directional emissivity, we use thin, subwavelength films of materials whose permittivity approaches and crosses zero, i.e. epsilon-near-zero (ENZ) behavior[42], as our model system. Planar ENZ films are known to support a leaky electromagnetic mode near the material resonance pole in the p polariztion, often called the Berreman mode[43-47]. This mode can couple to propagating free-space modes over a range of wave vectors and thus exhibits a directional response in its emissivity which can be controlled by adjusting the film thickness[43]. We begin by considering an ideal gradient ENZ framework consisting of finite number of thin films, each defined by a Drude-Lorentz permittivity function with the same loss factor ($\gamma$) but complementary resonance frequencies that vary gradually along the depth dimension. Through transfer matrix simulations we observe that the ideal gradient ENZ film exhibits broad-spectrum directional thermal emission where the angular extent of high emissivity (here, arbitrarily defined as $\epsilon > 0.8$) is $\Delta\theta = 30°$ and the angular selectivity (here, defined as $s = 1 - \Delta\theta/90°$) is $s = 0.67$, across a wide dimensionless operational range $\Delta\omega = 0.2$ from 1.5 to 1.7, described by the complementary resonance frequencies of the constituent thin films (Fig. S1b). In theory, by increasing the loss factor ($\gamma \times 2$) of the constituent thin films, the angular extent of high emissivity can be broadened ($\theta_{\gamma \times 2} = 5°$), and the angular selectivity can be reduced ($s = 0.61$) (Fig. S1c). Alternatively, by reducing the loss factor ($\gamma/2$) of the constituent thin films, one can narrow ($\theta_{\gamma/2} = 5°$) the angular extent of high emissivity, and the directional response of the thermal emitter can become sharper ($s = 0.72$), consistently across the entire bandwidth (Fig. S1a). This capability is in marked contrast to previously reported reconfigurable platforms employing cavity modes[29-32], where the spectral bandwidth of the directional emission was typically narrow bandwidth and the resonance frequency itself changed significantly as a function of the reconfigured directionality of high emissivity.

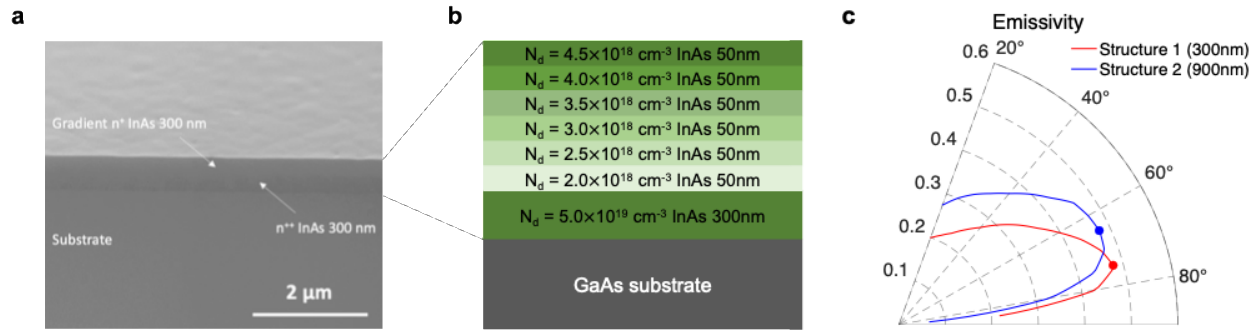

**Fig. 1 | Configuration of an InAs-based gradient ENZ thin film manifesting broadband directional thermal beaming.** Schematic of the InAs-based gradient ENZ structure with a doping concentration range from $2.0 \times 10^{18}$ cm$^{-3}$ to $4.5 \times 10^{18}$ cm$^{-3}$ and individual layer thickness of 50 nm (**a**). The SEM image of the experimentally fabricated multilayer InAs film structure, with labels identifying the materials used and the total thickness of the gradient ENZ layer (**b**). **c**, Directional response over a broad wavelength range of operation (from 12.5 to 15 μm) for structure 1 (300 nm) and 2 (900 nm), with peak average emissivity observed at 75° and 65°, respectively.

We showcase this capability using InAs-based gradient ENZ photonic structures. The first structure we designed and grew by molecular beam epitaxy was composed of InAs thin films of the same thickness with the doping concentration ranging from $2.0 \times 10^{18}$ cm$^{-3}$ to $4.5 \times 10^{18}$ cm$^{-3}$ (Fig. 1b). In the first structure, the thickness of the individual layers was 50 nm each and they were epitaxially grown atop a heavily doped InAs layer with a doping concentration of $5.0 \times 10^{19}$ cm$^{-3}$ and thickness of 300 nm. The gradient ENZ structure was grown on a 50 mm (2 inch) semi-insulating GaAs substrate, with a total InAs film thickness of 600 nm (Fig. 1a). We observe "thermal beaming" throughout a broad wavelength range from 12.5 to 15 $\mu$m, where the angular extent of p-polarized average emissivity above 0.4 throughout the entire bandwidth is $\Delta\theta = 20°$ centered at 75° (Fig. 1c and Fig. S2a).

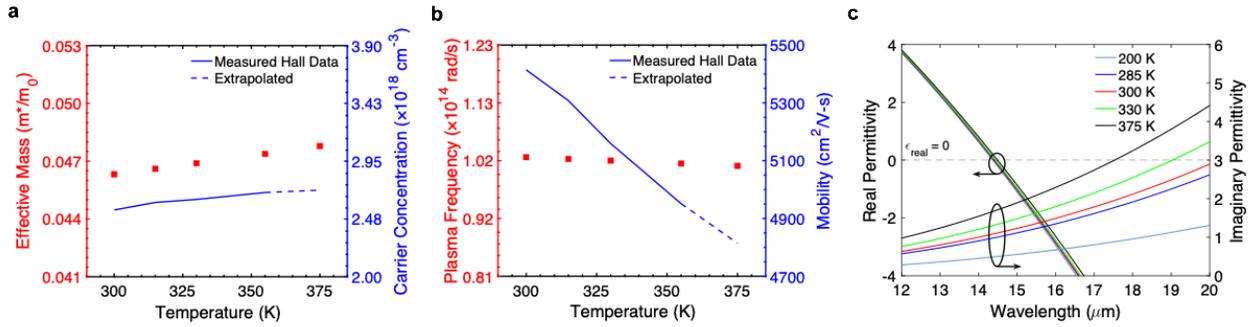

**Fig. 2 | Temperature-dependent Drude parameters of doped InAs.** Effective mass of the free-carriers calculated using the Kane model and experimentally measured Hall carrier concentration of a doped InAs thin film ($n$ =2.5×10$^{18}$ cm$^{-3}$ ) at different temperatures (**a**). The plasma frequency was calculated using Eq.1 (**b**), incorporating the values from (**a**). The measured Hall mobility shows a decreasing trend indicating the enhancement of scattering processes with increasing temperature (**b**). The real and imaginary parts of the permittivity at different temperatures (**c**), calculated based on the temperature dependent Drude parameters studied in (**a**) and (**b**). The invariance of the real part of the permittivity combined with the gradual tunability of the imaginary part of the permittivity makes gradient ENZ photonic structures based on doped InAs thin films ideal for a thermally reconfigurable metamaterials platform.

We first describe the temperature-dependent optical response of doped InAs, using the Drude formalism incorporating established theoretical and empirical models for the temperature-dependent bandgap, band curvature, and Fermi distribution:

$$\varepsilon = \varepsilon' + i\varepsilon'' = \varepsilon_\infty \left(1 - \frac{\omega_p^2[T]}{\omega^2 + i\omega\Gamma[T]}\right) \quad , \quad \omega_p^2[T] = \frac{n[T]e^2}{\varepsilon_\infty \varepsilon_0 m^*[T]} \qquad (1)$$

We study the temperature-dependence of the Drude parameters for a doped InAs thin film with a room temperature free-carrier concentration $n = 2.5 \times 10^{18}$ cm$^{-3}$, effective mass $m^* = 0.038 m_0$ and scattering rate $\Gamma$ = 5.6 THz[48]. The free-carrier concentration was chosen such that it corresponds to a value within the doping concentration range of the InAs-based gradient ENZ photonic structure. The effective mass of the free-carriers was calculated using the Kane model[49,50], accounting for the non-parabolicity of the conduction band-edge of InAs (see Fig. 2a and Supplementary Section 4). Here, the temperature-dependent band gap was modeled using the empirical Varshni model[51,52] for InAs (see Supplementary Section 4) and the free-carrier concentration was empirically determined by temperature-dependent Hall measurements using our reference InAs film epitaxially grown on a semi-insulating substrate (Fig. 2a). The temperature-dependent plasma frequency $\omega_p[T]$ was calculated using Eq. 1 (Fig. 2b), incorporating the non-parabolic effective mass $m^*[T]$ and the free-carrier density $n[T]$ from Fig. 2a. The plasma frequency remains relatively constant with increasing temperature, which can be explained by the increase of the electron effective mass balanced with the minimal increase in carrier density for a doped InAs thin film. The temperature-dependent scattering rate $\Gamma[T]$ was also obtained by curve-fitting experimentally reported values inferred from the literature (Table S1, Supporting Information). Although low frequency conductivity measurements are not necessarily a good indicator of optical conductivity[37], the temperature-dependent Hall mobility measurements clearly show a decreasing trend (Fig. 2b), indicating the enhancement of scattering processes with increasing temperature[53,54].

Based on the temperature dependent Drude parameters studied, we numerically calculate the real and imaginary parts of the permittivity of a doped InAs thin film ($n = 2.5 \times 10^{18}$ cm$^{-3}$ ) for different temperatures using Eq.1. With increasing temperature from 200 K to 375 K, the real part of the permittivity, hence, the ENZ wavelength (~14.5 $\mu$m) slightly red-shifts, however, stays relatively constant throughout this temperature range (Fig. 2c). The imaginary part of the permittivity at the ENZ wavelength (~14.5 $\mu$m) shows a gradual increase from 0.5 to 1.7 with increasing temperature from 200 K to 375 K. The key feature these measurements reveal, the thermal invariance of the real part of the permittivity combined with the gradual change in the imaginary part of the permittivity with temperature, makes this a unique platform. It enables us to thermally reconfigure the angular extent of high emissivity for a fixed spectral emissivity bandwidth.

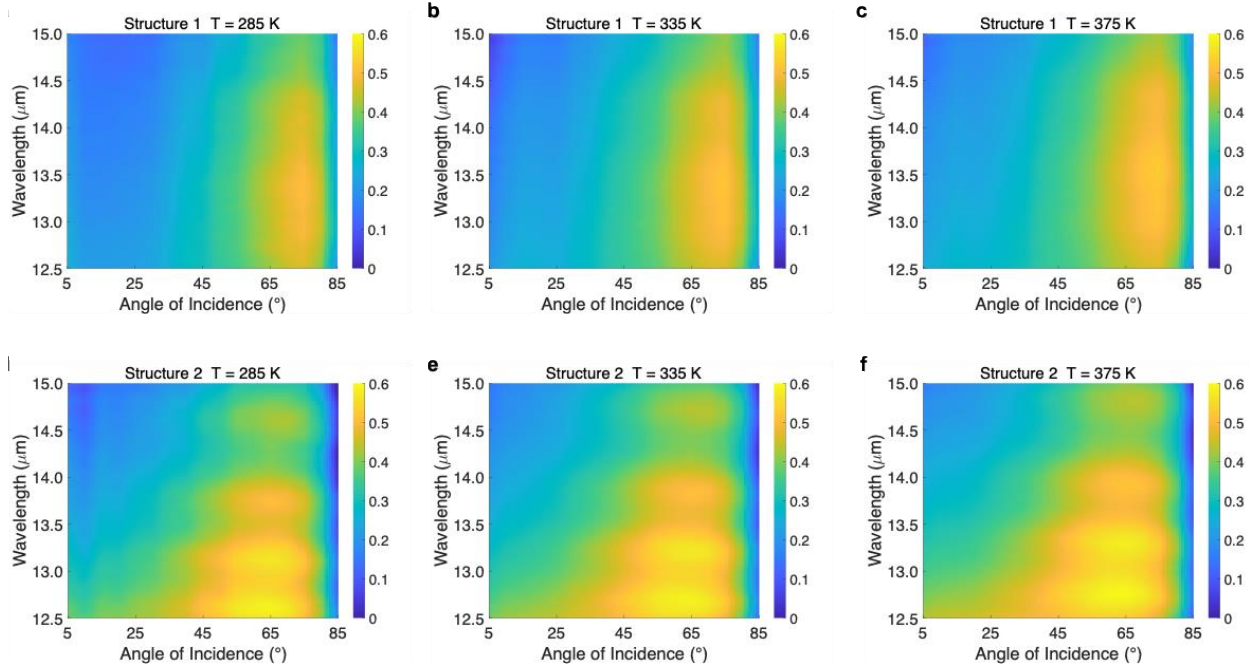

**Fig. 3 | Experimental tuning of infrared emisivity for two broadband directional thermal emitters. a-c** Measured emissivity spectra in the p polarization for structure 1 (300 nm) at (**a**) 285 K, (**b**) 335 K and (**c**) 375 K. Tthe structure's temperature is modulated by a small form factor thermoelectric Peltier module.. Measured emissivity spectra for structure 2 (900 nm) at (**d**) 285 K, (**e**) 335 K and (**f**) 375 K.

To experimentally verify the tunability of the angular extent of thermal beaming, we measure the emissivity spectra of the fabricated structure (structure 1) as a function of angle and wavelength in the p-polarization (Fig. 3a-c), using a FTIR spectrometer as the structure's temperature is modulated by a small form factor thermoelectric Peltier module. Since the gradient ENZ layer was grown on an optically thick metal layer, the transmissivity ($T$) is equal to 0 and the emissivity ($\epsilon$) is determined through the measured reflectivity ($R$) as a function of polar angle of incidence ($\epsilon = 1 - R$). The emissivity spectra exhibit strong directional emission, where a gradual increase in the angular extent of directional emissivity is observed with increasing temperature from 285 K to 335 K and 375 K. The spectral extent on the other hand, remains fixed across a broad wavelength range from 12.5 to 15 $\mu$m associated with the ENZ wavelength range of the constituent doped InAs thin films of the gradient ENZ layer[48].

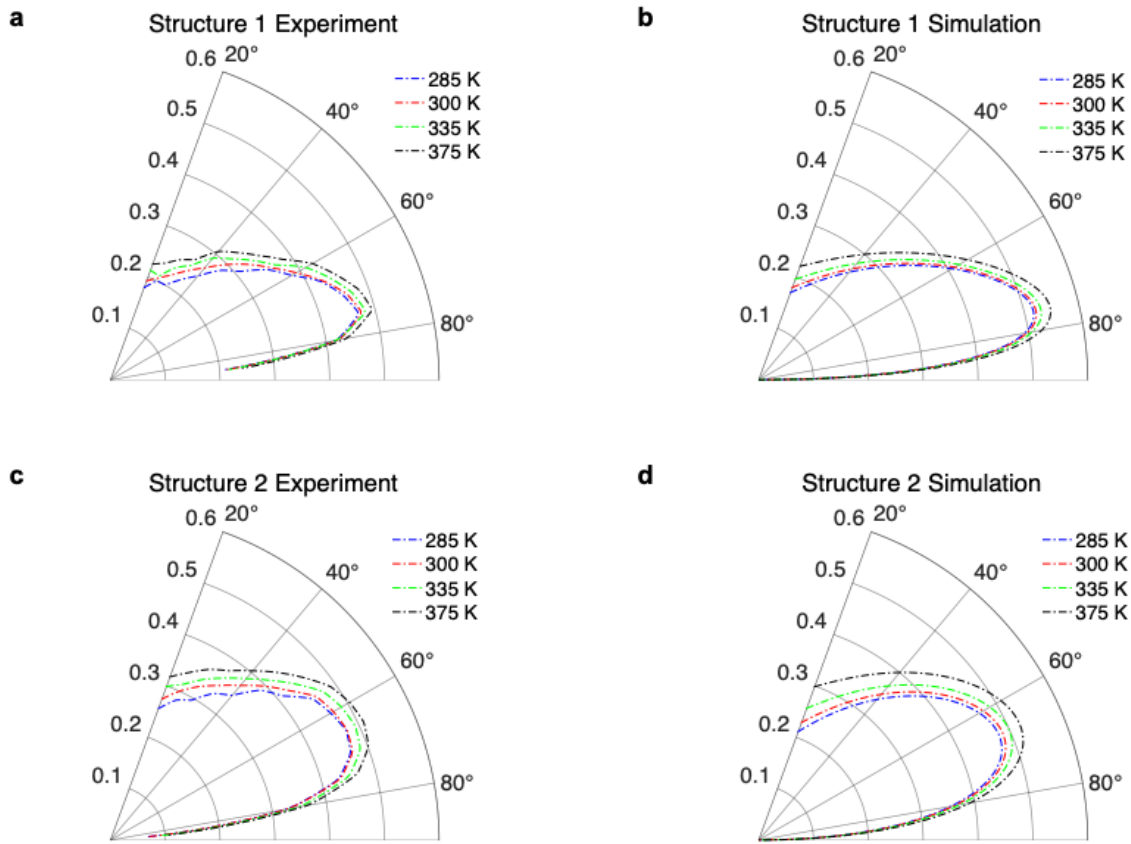

**Fig. 4 | Comparing experimental measurements with simulation results. a,c** Measured average p-polarized emissivity over a broad wavelength range of operation (from 12.5 to 15 μm) at different temperatures for structure 1 (**a**) and structure 2 (**c**). **b,d** Simulated average emissivity over the same operational range for structure1 (**b**) and structure2 (**d**). The simulation results show gradual tuning of the angular extent of directional thermal emission and match well with the experimental results.

To better characterize the thermal free-carrier effect on directional emissivity, we plot the overall p-polarized emissivity for the first structure over its wavelength range of operation (12.5 to 15 $\mu$m) in polar coordinates (Fig. 4a). As compared to the emissivity spectra at room temperature, we observe a narrowing of the angular width of high emissivity ($\epsilon > 0.4$) at 285 K, and a gradual increase of the angular width by 5°, corresponding to a gradual decrease in angular selectivity from 0.79 to 0.73, as the temperature increases from 285 K to 335 K and 375 K. Hence, we clearly see that one can gradually narrow or broaden the angular extent of the broadband beaming effect by controlling the temperature of the gradient ENZ photonic structure for an arbitrary, and fixed wavelength range of operation. To validate our experimental results, we simulated the average emissivity in the p-polarization of structure 1 over the wavelength range of operation (12.5 to 15 $\mu$m) in polar coordinates (Fig. 4b), using the transfer matrix method incorporating the temperature-dependent Drude formulism developed in Eq. 1. Although the measured emissivity is lower than what was predicted from the simulated data, the angles of operation as well as the gradual trend of the thermal reconfigurability with respect to temperature is in good agreement with the simulation. We note that the constraints in the absorptivity measurement setup did not allow for measurements at temperatures below 10°C. However, further simulations at lower temperatures (200 K) suggest that the angular extent of high emissivity could be further reduced by as much as 8° compared to the spectra at 375 K, exhibiting strong directional emission reaching an angular selectivity of 0.81, while maintaining its broad wavelength range of operation (Fig. S2a). This highlights the remarkable tunability of the directional response of broad-spectrum directional thermal emission of InAs-based gradient ENZ photonic structures by thermal free-carrier effects.

We fabricated a second semiconductor gradient ENZ structure (structure 2) with a doping concentration range of the gradient ENZ thin film ranging from $2.0\times10^{18}$ cm$^{-3}$ to $4.5\times10^{18}$ cm$^{-3}$, identical to structure 1, however, with a larger individual layer thickness of 150 nm, so that the total thickness of the gradient ENZ thin film was 900 nm; from which we expected to observe thermal tuning of the directional response at smaller angular ranges for the same temperature range (< 400 K). At room temperature, we observe strong directional emissivity throughout the same wavelength range from 12.5 to 15 $\mu$m, as in structure 1. However, the angular extent of the p-polarized average emissivity above 0.4 across the entire bandwidth was $\Delta\theta = 30°$, centered at a smaller angular range around 65° (Fig. 2c and Fig. S1b). This difference in the angular response of structure 2 originates fundamentally from the spatial shift of the optical mode supported by thicker Berreman films relative to the thinner ones in structure 1 (Fig. S3). The thermally reconfigured emissivity spectra exhibits, similar to structure 1, strong directional emission throughout a constant operational wavelength range from 12.5 to 15 $\mu$m, where a gradual increase in the angular extent of directional emissivity is observed with increasing temperature from 285 K to 335 K and 375 K (Fig. 3d-f). We emphasize that the broad spectral range of this directional modulation scheme can be arbitrarily prescribed by changing the doping concentration range of the gradient ENZ layer (Fig. S4). Upon a closer examination of the p-polarized average emissivity in polar coordinates, we observe a narrowing of the angular width of high emissivity ($\epsilon > 0.4$) at 285 K, however, as compared to structure 1, a larger overall modulation of the angular extent of high emissivity of 10°, corresponding to a gradual decrease in angular selectivity from 0.57 to 0.46, as the temperature increases from 285 K to 335 K and 375 K (Fig. 4c). As with structure 1, we compare the experimental results with simulations using the transfer matrix method and see excellent agreement, with both simulations and experiments exhibiting a larger 10° modulation of the angular width of high emissivity across the same temperature range (< 400 K), as compared to structure 1 (Fig. 4c,d). Additional simulations at lower temperatures (200 K) also suggest that the angular range of high emissivity could be further reduced as with structure 1, but as much as 15° compared to the spectra at 375 K, reaching an angular selectivity of 0.76 (Fig. S2b). A similar volumetric effect of thicker InAs layers exhibiting larger modulation relative to thinner ones has been reported in previous magneto-optic studies[33].

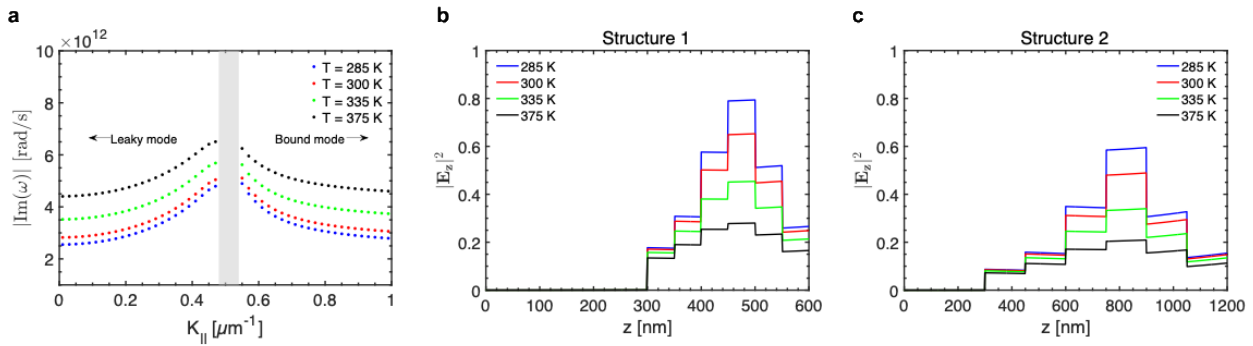

**Fig. 5 | Numerically calculated dispersion relation and electric field intensity distribution at different temperatures. a,** The calculated imaginary frequency in the dispersion relation of doped InAs Berreman films at different temperatures. The narrowing and broadening of the angular extent of thermal beaming is due to the modulation of the degree of attenuation of the electromagnetic wave as it oscillates in the medium. **b,c** The calculated electric field intensity distribution for the peak emissivity angle illumination at the center wavelength ($\lambda \approx 13.5 \ \mu m$) of the operational wavelength range for both structure 1 (**b**) and structure2 (**c**). With increasing temperature, we observe a reduction in the intensity of the localized field across the gradient ENZ layer, as the resonant optical modes are progressively converted into lattice vibrational modes.

To elucidate the physical origin of the temperature-dependent beaming effect, we numerically calculate the temperature-dependent dispersion relation of these leaky Berreman modes[43,44] (Fig. S5 and Fig. 5a). In Fig. 5a, we show the imaginary frequency in the dispersion relation of a doped InAs Berreman film at different temperatures. Here, we see that the imaginary frequency increases with increasing temperature from 285 K to 335 K and 375 K. From a fundamental optical mode point of view, the imaginary frequency corresponds to the degree of attenuation of the electromagnetic wave as it oscillates within the medium. This degree of attenuation can place limits on the directional contrast of emissivity. If the attenuation is larger for the constituent ENZ films at wavelengths near the ENZ wavelength, high emissivity will occur over a wide range of wave vectors, thus broadening the angular extent of thermal beaming. The real part of the frequency in the dispersion relation, on the other hand, remains constant with temperature modulation, indicating the invariance of the oscillation frequencies of these localized modes within these temperature ranges (Fig. S5). From a materials point of view, this thermal free-carrier effect could be understood as the material becoming more lossy and thus absorbing the incident optical power rather than allowing it to oscillate or propagate freely. To validate this, we calculated the electric field intensity distribution for the peak emissivity angle illumination at the center wavelength ($\lambda \approx 13.5\ \mu m$) of the operational wavelength range for both structure 1 and 2 (Fig. 5b,c). At room temperature, we observe a pronounced electric field enhancement at the center of the gradient ENZ layer, corresponding to the constituent layers supporting the Berreman mode at the central wavelengths within the high emissivity spectral bandwidth. Notably, the heavily doped InAs layer exhibits negligible electric field intensity in the simulations, remaining highly reflective over the wavelength range of operation and effectively pushing the field into the gradient ENZ layer. With decreasing temperature, however, we observe an enhancement in the intensity of the localized field. Conversely, with increasing temperature, we observe a reduction in the intensity of the localized field across the gradient ENZ layer, as the resonant optical modes are progressively converted into lattice vibrational modes. We also note that the overall spatial distribution of the electric field enhancement remains constant at the center of the gradient ENZ layer, indicating the invariance of the optical resonance frequency leading to the fixed broad-spectrum operational range of thermal beaming, regardless of the temperature.

**Conclusion**

We have proposed and experimentally demonstrated a tunable and reconfigurable photonic platform for directional emissivity using InAs-based gradient ENZ materials, where the angular extent of thermal beaming can be tuned via thermal free-carrier effects across a broad spectral emissivity bandwidth. We show a 5° increase in the angular extent of directional emissivity with temperature increases below 400 K, over a broad operational wavelength range from 12.5 to 15 µm, for the first structure we fabricated. For the second structure, we observe thermal reconfiguration of the directional response at smaller angular ranges within the same temperature range, but with a larger overall increase of the angular extent of high emissivity, reaching 10°. Our results show that the angular range of high emissivity can be further reduced by 8° and 15° at lower temperatures (200 K) compared to the spectra at 375 K for structures 1 and 2, respectively, where the directional response of the thermal emitter becomes sharper and more angularly selective across a prescribed bandwidth by thermal free-carrier effects. The reconfigurability of strong directional thermal emission to particular angular ranges over arbitrary broad spectral ranges, may yield substantial gains in device efficiency for existing technologies[55]. In particular, we believe that the III-V semiconductor-based nature of our thermally reconfigurable emissivity platform suggests its potential to inspire novel integrated photonic device concepts in the infrared, where the tunability of angular selectivity in broad-spectrum directional thermal emission can be leveraged.

**Acknowledgements**

This material was based upon work supported by the National Science Foundation under grant no. ECCS-2146577